# Threats and Limitations of Terrestrial Broadcast Attacks

Benjamin Michéle, Iván Peña, *Member, IEEE*, and Pablo Angueira, *Senior Member, IEEE*

*Abstract*—The DVB standard does not mandate the use of authentication and integrity protection for transport streams. This allows malicious third parties to replace legitimate broadcasts by overpowering terrestrial transmissions. The rogue signal can then deliver a malicious broadcast stream to exploit security vulnerabilities on Smart TVs (STVs) in range. We implemented a proof-of-concept attack based on a malicious Hybrid Broadcast Broadband TV app, able to acquire permanent system-level access to an STV over the air, in less than 10 s. These attacks, however, are severely limited in range due to required co-channel protection ratios (CCPRs), which is in direct contradiction to previous publications. We present evidence for these limitations in form of laboratory experiments, extensive simulations, and field measurements. To this end, we developed an automated, low-cost method for CCPR determination, as well as a method for non-disruptive attack range measurements based on a gap filler and the resulting channel impulse response.

## I. Introduction

During the last decade, many countries have completed the transition from analog to digital television (DTV). Now, more than a billion households have access to DTV services [1]. Broadcasters in Europe and around the world employ the Digital Video Broadcasting (DVB) standard. DTV has also enabled the introduction of interactive TV services. One of these services is Hybrid Broadcast Broadband TV (HbbTV) [2], which has been adopted in 25 countries. HbbTV builds upon existing Web standards, which allows vendors to leverage existing code to implement runtimes and developers to rapidly roll out applications. Current Smart TVs (STV) have built-in support for HbbTV, which has led to high acceptance among consumers.

The DVB and HbbTV standards, however, do not mandate the use of authentication and integrity protection for transport streams and applications, respectively. This allows malicious

This work was supported by the Spanish Ministry of Economy and Competitiveness under Project 5G-newBROS (TEC2015-66153-P MINECO/FEDER, UE). *(Corresponding author: Benjamin Michéle.)*

B. Michéle is with Technische Universität Berlin, 10587 Berlin, Germany (e-mail: benjamin.michele@tu-berlin.de).

I. Peña and P. Angueira are with the University of the Basque Country, 48013 Bilbao, Spain (e-mail: ivan.pena@ehu.eus; pablo.angueira@ehu.eus).

third parties to impersonate broadcast stations, which in turn can be abused to deliver and launch arbitrary HbbTV apps. A malicious HbbTV app can exploit (vendor-specific) security vulnerabilities on the STV and thereby permanently gain full control over the device. STVs have become an attractive target due to powerful hardware, connection to (private) local networks and the Internet, and their widespread use. To demonstrate the feasibility of such attacks, we implemented a fully functional proof-of-concept (PoC) attack. The PoC overpowers a regular DVB-T broadcast with a signal that contains a malicious HbbTV app. This app is automatically started on the STV and then gains full control by exploiting a vulnerability in the STV's media playback system. The attack and subsequent infection do not require any user interaction and are invisible to victims.

There are, however, significant limitations to broadcast-originated attacks, both in terms of range and stealthiness. Previous publications [3], [4] have calculated the attack range solely based on (free-space) path loss, ignoring required co-channel protection ratios (CCPRs). Taking into account the CCPR, however, results in a *significantly* reduced attack range. Furthermore, the resulting co-channel interference (CCI) creates a huge mush area, in which neither signal can be received. These customers—cut off from broadcast service—are likely to file complaints, which significantly limits the presumed stealthiness of such an attack.

This work presents range and mush zone estimations that take into account required CCPRs, along with extensive simulations for an area in Los Angeles. These results have been validated both in the lab and in field experiments. We also present an automated, low-cost method to measure CCPRs on modern STVs together with the obtained results. Furthermore, this work presents a novel method to determine attack ranges in the field, without disrupting the regular broadcast delivery.

The study's findings are, in principle, applicable to all unauthenticated terrestrial broadcast systems that convey a trigger in the broadcast that causes a STV to access an Internet resource, such as ATSC 2.0 in the U.S. [5], BML Datacasting in Japan [6], and DMB in China [7]. DVB and HbbTV were chosen as the subjects of the study due to their global importance and easy access to tools, equipment, and signals.

The rest of the paper is organized as follows. Section II provides the necessary background for the following sections. Section III describes the attack surface of a modern Smart TV. Section IV presents our broadcast-assisted attack and the PoC implementation. Section V explores the limitations to our attack, both in terms of range and detectability. Section VI provides the results of our attack simulations. Section VII introduces a novel approach to determine attack ranges in the field based on a gap filler and the resulting channel impulse response. Finally, Section VIII closes with the conclusion.

## II. Background

This section provides the background on the technologies and previous publications related to this work.

### A. HbbTV

Hybrid Broadcast Broadband Television (HbbTV) [2] allows broadcasters to augment their TV programs with interactive applications. HbbTV includes information on available applications in the DVB transport stream. Every program can have several associated applications, one of which is started automatically upon selection of the respective program. In general, this so-called *red button* application is supposed to indicate the presence of an HbbTV application by displaying a custom red button in the lower right-hand corner. If the user presses the red button on the remote control, the application will present its user interface, e.g., an electronic program guide (EPG) or catch-up TV.

*1) Signaling:* The broadcast transport stream contains a Program Map Table (PMT), which indicates to the receiver the available programs and their respective components. One of these components is the Application Information Table (AIT), which lists all available applications for the respective program. Furthermore, it indicates which of these applications should be started automatically and which transports can be used to fetch the applications.

*2) Transports:* There are two transports available for the delivery of HbbTV apps: HTTP(S) and the DSM-CC object carousel [8]. In the former case, applications are hosted on the broadcast station's Web server and fetched via HTTP(S) by the receiver. The alternative is to embed the application in the broadcast stream itself, i.e., in an object carousel. This allows broadcasters to support receivers that are not connected to the Internet, at the cost of increased broadcast link capacity requirements.

*3) Version:* Currently deployed devices comply to HbbTV version 1.0 and 1.5 from 2010 and 2012, respectively. The specifications for HbbTV 2.0 and 2.0.1 [2] were released in 2015 and 2016, respectively; compliant devices are expected for 2017. HbbTV 2.0 features several improvements, e.g., full HTML5 and companion screen support, push VOD, and enhanced privacy options regarding the use of HTTP cookies. HbbTV 2.0.1 adds features required for deployment on receivers in the U.K. and Italy. The security concerns raised by our work, however, are not addressed by these updates; this is planned for a subsequent release [9] (see Section IV-B5). The experiments presented in this paper were performed on devices compliant to HbbTV 1.x, but the results apply to all versions.

### B. Related Work

Smart TVs have been repeatedly shown to contain exploitable security vulnerabilities, resulting in complete control over the device. Most of the attacks require the attacker to have physical access to the device, or to the local network [10]. Michele and Karpow [11], however, demonstrated that STVs' integrated media player could be attacked with malicious media files played back from attached USB sticks. The same attack was used to extend the attack surface to Web browsers, apps, and HbbTV—all of these components make use of the same vulnerable media playback system [12].

Oren and Keromytis [3], [4] proposed a series of request forgery attacks using HbbTV apps delivered via the terrestrial broadcast channel. With regard to the physical properties of the broadcast channel and the resulting attack range, however, the work is purely theoretical. They did not overpower an existing channel but instead connected a modulator directly to the STV. The presented range calculations thus ignore required CCPRs, leading to an overestimated attack coverage by orders of magnitude. This is explained in detail in Section V.

## III. Smart TV Attack Surface

Virtually all Smart TVs employ a Linux-based operating system. Many STV models implement most of the functionality in a single proprietary application with unrestricted system privileges. This application runs multiple threads to provide the user interface, apps, media playback, etc.; much of which is implemented in open source libraries from third parties.

Software projects of this size are likely to contain exploitable vulnerabilities. If they are not patched by the vendor, they can be abused by attackers to gain unauthorized access to STVs and subsequently to connected networks and devices. The current STV ecosystem, however, is particularly ill-suited for guaranteeing a secure software stack.

First and foremost, the STV's system software—the firmware—cannot be updated incrementally; instead, up to several-hundred megabytes have to be downloaded and installed for every update, however small. This severely limits the rate at which firmware updates can be deployed, especially in terms of user acceptance. Furthermore, STV vendors introduce a variety of devices each and every year, which makes it increasingly difficult to supply current and old models with (timely) updates.

In combination with open source software (OSS), this becomes a real problem. STV vendors have to rigorously monitor security announcements for all of the libraries in use on current and previous STV models. Any newly discovered vulnerability entails a time-consuming firmware development, testing, and deployment process. As a result, legacy STV models are cut off from updates, and current models suffer from delayed updates leading to potentially vulnerable devices.

Additional bugs can be identified in proprietary software by reverse engineering or fuzzing. Attackers can target all of these vulnerabilities. Their exploitation is comparatively easy, as many STVs lack security best practices established in the PC world. Privilege separation, sandboxing, and exploitation countermeasures such as Data Execution Prevention (DEP) and Address Space Layout Randomization (ASLR) [14] have only recently begun to see a more widespread deployment. An attacker controlling the broadcast channel can exploit vulnerabilities in the HbbTV runtime, the media playback, and the components related to broadcast processing.

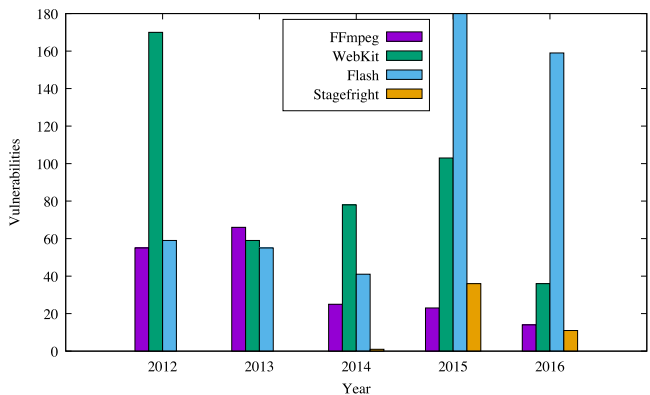

Fig. 1. Number of vulnerabilities published in libraries used on STVs in the past five years, based on CVE entries (FFmpeg: all entries; Apple WebKit, Adobe Flash, Android Stagefright: only code execution entries) [13].

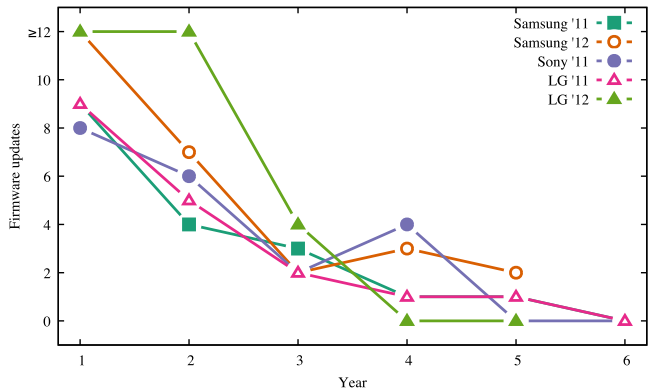

Fig. 2. Number of firmware updates issued per year for five different STV models from 2011 and 2012.

### A. HbbTV Runtime

HbbTV apps are written in CE-HTML (or HTML5 for HbbTV 2.x) and therefore most runtimes are based on existing Web browser projects. A popular example is WebKit, which is used by several STV vendors to implement HbbTV. New vulnerabilities in Web runtimes, however, are discovered and published regularly, leading to frequent updates on the PC platform. Fig. 1 shows the number of code execution vulnerabilities reported for WebKit during the past five years. On STVs, however, these updates have to be incorporated into the firmware, leading to substantial delays. In the case of legacy devices, the frequency at which firmware updates are developed decreases and eventually stops completely. Fig. 2 illustrates this with devices from three popular STV vendors: After a period of five years, firmware updates essentially cease to be published.

### B. Media Playback

STVs employ a central component for media retrieval, analysis, decoding, and display—the media playback system. This system is used by other STV components such as apps, the HbbTV runtime, and the built-in media player. Several STV vendors use open source libraries to implement the media handling, e.g., FFmpeg, GStreamer, or libstagefright [15]. Handling media files, however, is a complex task, which has resulted in the continuous discovery of vulnerabilities (see Fig. 1). These vulnerabilities can be exploited on STVs at large scale, as we have previously demonstrated for STVs employing FFmpeg [11], [12]. Further vulnerabilities have recently been discovered in libstagefright, the media playback system of Android devices [16], [17], and in GStreamer [18]. Table I lists open source media processing frameworks used on several popular STVs.

TABLE I
Media Playback Libraries Used on Smart TVs

| Component | Version | Released | Vendor | Model |
|---|---|---|---|---|
| FFmpeg | SVN-r158** | 11.2008 | A | 2009 |
| | SVN-r19089 | 05.06.2009 | | 2010 |
| | | | | 2011 |
| | 0.6.90-rc | 03.04.2011 | | 2012 |
| | | | | 2013 |
| | | | | 2014 |
| | n1.0 | 28.09.2012 | | 2015 |
| | SVN-r17783 | 03.03.2009 | B | 2013 |
| | 0.6.1 | 18.10.2010 | C | 2014 |
| | n1.1.1 | 20.01.2013 | D | 2014 |
| GStreamer | 0.10.36 | 20.02.2012 | B | 2013 |
| | | | C | 2014 |
| | | | A | 2015 |
| Stagefright | 1.2 | 12.02.2013 | D | 2014 |

### C. Service Information

DVB defines a number of service information (SI) tables [19], which are delivered in the broadcast transport stream and are processed by the Smart TV. If the processing software contains vulnerabilities, this may be exploited by transmitting maliciously crafted tables. Vulnerable implementations of the DVB subtitling system [20] could be attacked similarly. With the ongoing consolidation of STV hardware platforms and associated software stacks, a single vulnerability in this component can potentially affect several STV vendors and a wide variety of devices.

## IV. Broadcast-Assisted Attacks

The DVB standards do not require broadcasts to be authenticated. As a result, receivers are unable to distinguish between authentic and forged broadcasts. Malicious third parties can abuse this to alter the received programming.

In the past, however, this was not considered a serious threat, as the potential damage was limited to disseminating fake alerts, advertisements, etc. However, this has changed with the introduction of Smart TVs and interactive digital services such as HbbTV. An attacker controlling the broadcast channel is able to deliver malicious applications to receivers, in particular STVs. Combined with vulnerable system software on STVs, this can become a real threat for STV owners.

Compromising a STV consists of two main parts. First, the broadcast channel has to be controlled, i.e., for terrestrial broadcasts, victim STVs must receive the rogue signal instead of the regular signal. Second, the rogue signal must contain malicious code to take over victim STVs, e.g., a malicious HbbTV application (see Section III). Limitations to this attack are discussed in detail in Section V. This section is based on our previous work [12].

### A. Rogue DVB-T Signal

An attacker has to overpower the regular DVB-T or DVB-T2 signal arriving at the receiver; both are vulnerable and we will thus refer to them as DVB-T. Several attack scenarios are conceivable: The targeted, the untargeted small range, and the regional repeater scenario. In the targeted scenario, the STV of a specific person of interest is compromised, using a highly directional antenna. For the small range scenario, all TVs in the vicinity of the attacker's transmitter are the target.

In some ways, this attack can be compared to man-in-the-middle (MITM) attacks in GSM networks [21]. The GSM specification requires that mobile stations (MS) authenticate to the network, but not vice versa. In DVB-T, broadcast stations do not authenticate to receivers, either. The use of allocated frequencies, however, differs between the two systems. In GSM, the base stations (BS) in the vicinity of the MS operate on different frequency bands and the MS connects to the BS with the strongest signal. An attacker can therefore operate a rogue BS on an *unused* frequency, i.e., there is no need to physically overpower another BS on the same channel. DVB-T receivers, on the other hand, remain tuned to the frequency of the currently selected channel, thus forcing an attacker to overpower the regular broadcast signal. As opposed to MITM attacks in the GSM network, this creates an area without proper reception of either signal, which we refer to as the *mush* zone (see Section V-A).

Another interesting target for attackers is broadcast relay stations or gap fillers [22]. Here, an attacker overpowers the regular broadcast signal at the receiving antenna of the retransmission station, provided the signal is picked up off-air. This allows the attacker to abuse the amplification service of the retransmission station and hence reach a large region.

Apart from DVB-T, a resourceful attacker can attempt to overpower a satellite up- or downlink, both of which has occurred repeatedly in the past [23]. Cable television systems can also be attacked, by injecting malicious signals into distribution lines. Even worse, an attacker could target cable TV headend facilities that are fed by satellite or terrestrial signals. The attacker-controlled signal is then distributed through the headend to a large amount of cable subscribers.

### B. Proof of Concept

Our proof-of-concept (PoC) implementation demonstrates that broadcast-assisted attacks are feasible in practice and can pose a threat to consumers, which helps vendors and standard bodies to understand the risks and develop countermeasures. Furthermore, being able to analyze an attack in practice gives us the opportunity to identify weak spots, both on the side of the target and the attacker, and design efficient countermeasures. It also allows us to verify to what degree previously proposed attacks are feasible in practice. Figure 3 illustrates

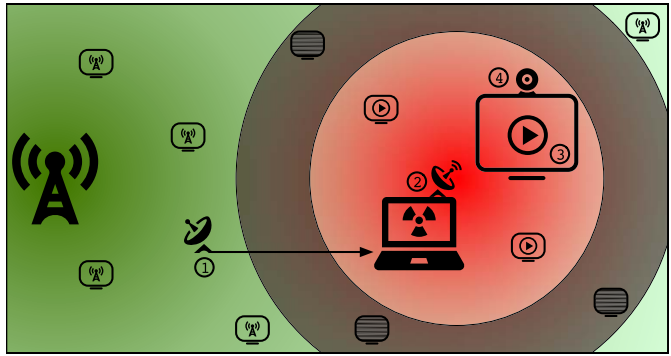

Fig. 3. Regular broadcast is picked up (1), modified to include malicious HbbTV app and retransmitted (2), invokes compromising media playback on target STVs (3), and finally executes payload, e.g., camera tapping (4). Note the shaded mush area, where neither signal has sufficient co-channel protection to be received properly (not to scale, see Fig. 7) [12].

```
1  # tune to station
2  tzap -r "Das_Erste" &
3  # grab and save raw transport stream
4  dvbsnoop -b -s ts -tsraw > reg.ts &
5  # replace ait and object carousel
6  tsmodder reg.ts +2070 ait.ts +2171 oc.ts > rogue.ts
```

Listing 1. Signal acquisition and modification [12].

the MITM attack scenario that was chosen for the PoC implementation. The regular terrestrial broadcast signal is picked up, modified to include a malicious HbbTV app, and retransmitted. STVs in range receive this signal and automatically execute the HbbTV app, thereby infecting themselves with persistent malware. It is important to note that no user interaction is required and the STV compromise is invisible to the victims.

*1) Regular Signal Acquisition:* Retransmission of the current program is required for a stealthy—invisible to the user—attack. The transport stream is acquired in real-time with a cheap USB-based, consumer-grade DVB-T receiver. Standard Linux tools are used to tune the receiver to the target station and acquire the raw transport stream, as shown in Listing 1.

*2) Transport Stream Modification:* The transport stream is modified to automatically launch the malicious HbbTV app on victim STVs. This does not require any user interaction, i.e., the user does not have to change the channel or stop a currently active HbbTV app. The HbbTV specification states that a running app can be killed via the broadcast signal by removing the app ID from the AIT. The AIT is thus modified to include both the original and the added malicious app with a new ID. In addition, the type is set to `PRESENT` and `AUTOSTART` for the original and malicious app, respectively. The STV thus kills the running app and immediately starts the malicious app. These modifications were performed with the open source tool collection OpenCaster from Avalpa [24] (see Listing 1).

*a) Transmission:* A DVB-T transmitter is required to retransmit the modified transport stream. Baseband processing can be performed either in hardware or software; the latter is known as Software-Defined Radio (SDR). The professional-grade DekTec DTU-215 and the cheaper Hides UT-100C use

| Vendor | Model | Type | Power [dBm] | Price [$] |
|---|---|---|---|---|
| DekTec | DTU-215 | DVB-T/C | (-15) | 1,600 |
| Hides | UT-100C | DVB-T | 8 | 150 |
| G.S.G. | HackRF | SDR | -7 | 300 |
| Ettus | USRP B210 | SDR | 10.5 | 1,100 |

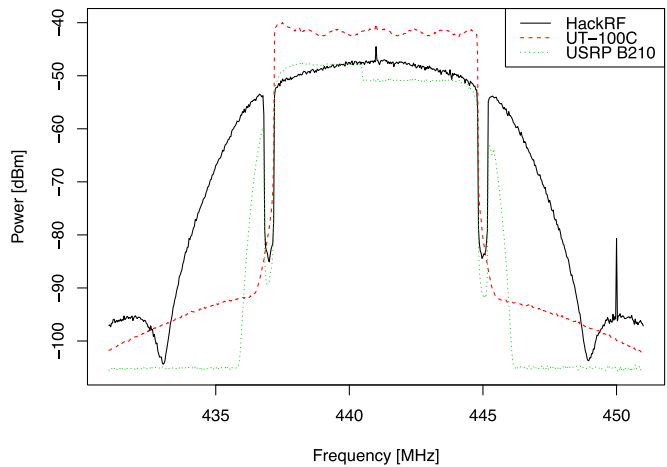

Fig. 4. Transmitter spectrum [12].

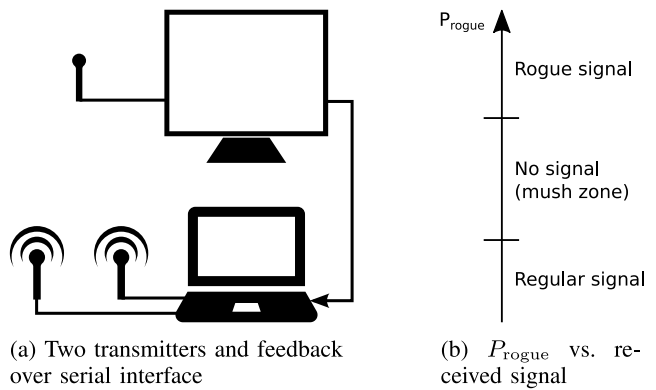

(a) Two transmitters and feedback over serial interface

(b) $P_{rogue}$ vs. received signal

Fig. 5. CCPR measurement setup [12].

the former, whereas the Great Scott Gadgets open source HackRF and the professional-grade Ettus Research USRP B210 use SDR. Measured signal strength, adjacent channel protection, and overall signal quality are provided in Table II and Fig. 4. We used the cost-efficient UT-100C for the majority of the experiments.

*b) Isolation:* A MITM attack requires a setup that is able to receive, modify, and retransmit the regular signal in real-time. The required level of isolation between the receiving and transmitting antenna is difficult to achieve without expensive professional equipment. In practice, however, there are several ways to solve this problem. An attacker can receive the regular signal outside of the mush zone and then use an out-of-band mechanism, e.g., the Internet, to deliver the broadcast stream to the rogue transmitter. Alternatively, if the station is broadcasting publicly available material such as an old movie, the attacker does not need to operate a receiver. Finally, the attack can be launched during commercial breaks, where the regular broadcast can be replaced with arbitrary commercials.

*3) Malicious App:* An HbbTV app is delivered either via HTTP(S) or the DSM-CC object carousel (see Section II-A2). Experiments with our PoC attack show that the HTTP(S)-based delivery is generally faster and that it is the only method to allow for media playback exploits. The total time required for an attacker to be on air to compromise a STV is *less than ten seconds*.

*4) Exploitation:* Our PoC attack exploits a vulnerability in the media playback system (see Section III-B). A malicious video file is included in the HbbTV autostart app, which is automatically started when the (invisible) app launches on the STV. The vulnerability is triggered upon playback, which results in a root shell on the STV. The video itself is never actually displayed and therefore does not interrupt the currently running broadcast. The final step after having gained access to the STV is the execution of malicious payload. Our PoC exploit establishes a connection to a remote server, which is used to load and run arbitrary code on the STV. For demonstration purposes, our code taps into the STV's camera and microphone and transmits the recorded stream over the Internet in real-time.

*5) Disclosure and Reaction:* We initially contacted the HbbTV association in June 2014 to responsibly disclose our findings on HbbTV-assisted attacks. We renewed this offer in August, which led to a live demonstration on a real device to the HbbTV Chairman in September. We have since been invited to demonstrate and discuss our results and possible countermeasures at various meetings of the HbbTV group, and also the DVB Technical Module (DVB TM) in January 2015. This collaboration has raised a thorough awareness for HbbTV-assisted attacks within the HbbTV and DVB group, and has created an active process to specify future protection measures. In May 2015, the HbbTV group announced [9] that they were working closely with the DVB group and that they had decided on an approach based on the authentication of critical MPEG-2 sections. An updated specification [25] protecting against the HbbTV-assisted malware delivery presented in this work is expected to be completed in early 2017.

## V. LIMITATIONS

The presented attack is subject to a number of limitations, most notably in terms of range and stealthiness. This section is the extended version of our work published in [12, Sec. 3.7], expanding the measurements for DVB-T and adding results for two DVB-T2 modes in active use.

### A. Co-Channel Protection Ratio

The attack scenario consists of a regular terrestrial broadcast and a rogue transmitter broadcasting on the same channel, thus overpowering the signal in its vicinity. In this co-channel situation, only the stronger of the two signals can be decoded while the other appears as co-channel interference (CCI). More precisely, the stronger signal can be decoded if its power ratio

TABLE III
CCPR Measured According to the Method Presented in Section V-A1 (M2) and Minimum Required C/N Measured With Professional Equipment (M1) vs. Calculated C/N Values From Reimers [26, Table 11.7] (Reim.) and ETSI [27, Table A.1] vs. CCPR by ITU-R [28, Table 15]. Values Are Given in dB for Gaussian, Ricean, and Rayleigh Propagation Channels

| Modulation | CR | Gauss | | | | | Rice | | | | Rayleigh | | | |
|---|---|---|---|---|---|---|---|---|---|---|---|---|---|---|
| | | M1 | M2 | Reim. | ETSI | ITU | M1 | Reim. | ETSI | ITU | M1 | Reim. | ETSI | ITU |
| QPSK | 1/2 | 2.0 | 2.5 | 3.1 | 3.5 | 5 | 2.8 | 3.6 | 4.1 | 6 | 4.1 | 5.4 | 5.9 | 8 |
| | 2/3 | 3.8 | 3.5 | 4.9 | 5.3 | 7 | 4.8 | 5.7 | 6.1 | 8 | 7.1 | 8.4 | 9.6 | 11 |
| | 3/4 | 4.7 | 5.0 | 5.9 | 6.3 | – | 5.9 | 6.8 | 7.2 | – | 9.1 | 10.7 | 12.4 | – |
| | 5/6 | 5.8 | 6.0 | 6.9 | 7.3 | – | 7.3 | 8.0 | 8.5 | – | 12.0 | 13.1 | 15.6 | – |
| | 7/8 | 6.4 | 7.0 | 7.7 | 7.9 | – | 8.0 | 8.7 | 9.2 | – | 13.9 | 16.3 | 17.5 | – |
| 16-QAM | 1/2 | 7.3 | 8.0 | 8.8 | 9.3 | 10 | 8.1 | 9.6 | 9.8 | 11 | 9.4 | 11.2 | 11.8 | 13 |
| | 2/3 | 9.6 | 10.0 | 11.1 | 11.4 | 13 | 10.6 | 11.6 | 12.1 | 14 | 12.7 | 14.2 | 15.3 | 16 |
| | 3/4 | 10.8 | 11.0 | 12.5 | 12.6 | 14 | 12.0 | 13.0 | 13.4 | 15 | 14.7 | 16.7 | 18.1 | 18 |
| | 5/6 | 12.1 | 12.5 | 13.5 | 13.8 | – | 13.4 | 14.4 | 14.8 | – | 17.5 | 19.3 | 21.3 | – |
| | 7/8 | 12.8 | 13.0 | 13.9 | 14.4 | – | 14.3 | 15.0 | 15.7 | – | 19.5 | 22.8 | 23.6 | – |
| 64-QAM | 1/2 | 11.5 | 11.5 | 14.4 | 13.8 | 16 | 12.4 | 14.7 | 14.3 | 17 | 13.9 | 16.0 | 16.4 | 19 |
| | 2/3 | 14.7 | 15.0 | 16.5 | 16.7 | 19 | 15.5 | 17.1 | 17.3 | 20 | 17.4 | 19.3 | 20.3 | 23 |
| | 3/4 | 16.2 | 16.5 | 18.0 | 18.2 | 20 | 17.3 | 18.6 | 18.9 | 21 | 19.8 | 21.7 | 23.0 | 25 |
| | 5/6 | 17.7 | 18.3 | 19.3 | 19.4 | – | 18.9 | 20.0 | 20.4 | – | 22.5 | 25.3 | 26.2 | – |
| | 7/8 | 18.6 | 19.5 | 20.1 | 20.2 | – | 19.9 | 21.0 | 21.3 | – | 24.7 | 27.9 | 28.6 | – |

w.r.t. the weaker signal exceeds a certain threshold — the co-channel protection ratio (CCPR).

The required CCPR depends on the transmission's modulation type and forward error correction (FEC), given as code rate (CR). If the actual ratio drops below the CCPR, neither signal will be decoded and the TV screen turns dark. The attack therefore creates three distinct areas of signal coverage: First, the area surrounding the rogue transmitter, in which the rogue signal strength exceeds the required CCPR w.r.t. the regular broadcast signal. Surrounding this area is a second area, in which neither signal is strong enough to achieve the required CCPR and TV screens thus turn dark. We will call this the *mush* area, adopting a term used in mediumwave broadcasting to describe areas with echo-like interferences. And finally the remaining area serviced by the regular broadcast signal, exceeding the required CCPR w.r.t. the rogue signal.

*1) Measurements:* Broadcasters aim to provide high-quality service and thus target C/N ratios that allow for "Quasi Error Free" (QEF) transmissions [22]. The C/N ratio required for an attack, however, is lower: Victim receivers must lock on to the signal for at least ten seconds, which is enough to decode the AIT and launch the malicious application. This is the evaluation criterion we chose for measuring CCPRs related to the attack scenario.

To determine the CCPR for all (8 MHz channel) DVB-T variants, we implemented an automated measurement setup as illustrated in Fig. 5a. It consists of two low-cost transmitters, a target TV, and a laptop controlling the measurements and reading serial output from the TV—alternatively, HbbTV requests can be used for the feedback channel if no serial output is available. Both of the transmitters use the same settings for modulation and CR. They are placed close to the STV's antenna and have a direct line-of-sight connection, resulting in an additive white Gaussian noise (AWGN) propagation channel. One of the transmitters simulates the regular broadcast signal and transmits with a low, fixed power. The second transmitter serves as the rogue station, starting with sufficiently higher power output, which is subsequently decreased by 1 dB in each step. During each step, the TV's serial output is monitored for changes in the signal lock status of its receiver. The measurement setup is able to determine CCPR values by associating the relation between the regular and rogue station's power and the three states mentioned above: Rogue broadcast, mush zone, and regular broadcast. If the receiver is able to maintain a steady signal lock for a period of ten seconds, the current CCPR is deemed sufficient to launch a successful attack; otherwise, an attack with this CCPR will result in mush zone. Figure 5b illustrates the signals decoded by the STV depending on the rogue transmitter power $P_{rogue}$. All measurements were repeated ten times and cross-checked on a different STV model. Neither the guard interval (GI) nor the transmission mode (2K or 8K) had a significant influence on the measured CCPRs. The resulting CCPR values are given in Table III as M2.

TABLE IV
Min. Required C/N [dB] for DVB-T2 Variants G2 (16Ke 64QAM CR3/5 PP2) and G8 (32Ke 64QAM CR2/3 PP4), Measured (M1) and Simulated (DTVP) [29]

| DVB-T2 variant [29] | Gauss | | Rice | | Rayleigh | |
|---|---|---|---|---|---|---|
| | M1 | DTVP | M1 | DTVP | M1 | DTVP |
| G2 | 12.8 | 14.8 | 13.3 | 15.1 | 15.2 | 16.9 |
| G8 | 14.1 | 15.7 | 14.7 | 16.1 | 16.8 | 17.9 |

For comparison, we measured the minimum required C/N ratio conforming to the Subjective Failure Point (SFP) assessment method [28]. A Rohde & Schwarz SFU was used to generate the broadcast signal, add noise, and apply the channel simulation. The signal was fed to a 2015 STV model, the Samsung UE40JU6450, which was used to assess the picture quality. The results are labeled as M1 in Table III and Table IV. The latter provides results for two DVB-T2 variants in active use by German broadcasters, labeled G2 (16Ke, 64-QAM, CR 3/5, GI 19/128, PP2, LDPC 64800, SISO) and G8 (32Ke, 64-QAM, CR 2/3, GI 1/16, PP4, LDPC 64800,

SISO) [29]. Additional values obtained from simulations are provided for DVB-T2 by DTVP [29, Annex A].

Reimers [26, Table 11.7] and ETSI [27, Table A.1] provide carrier-to-noise (C/N) ratios required to achieve a quasi-error-free (QEF) DVB-T transmission, i.e., less than one uncorrected error event per hour [22]. ITU-R provides measured CCPR values for a wanted DVB-T signal that is interfered with by another DVB-T signal [28, Table 15]. The values are chosen such that a QEF transmission is achieved, including a typical implementation margin.

Compared to our measurements, the values from ETSI and ITU are roughly 1 dB and 3 dB higher, respectively; the values calculated by Reimers are slightly higher than the values measured by us and slightly lower than the values given by ETSI (with the exception of 64-QAM and CR 1/2). Reimers, ETSI, and ITU target QEF transmissions, whereas our measurements reflect the minimum CCPR required to launch a successful attack.

*2) Attack Range and Controlled Area:* The Log-Distance path loss model [30] reflects that the average received signal power decreases logarithmically with the distance. This is accounted for by a path loss exponent (PLE) $n$ that describes the rate at which the signal power decreases. In free space, for instance, $n$ equals 2, whereas obstructions will increase the value of $n$. The resulting large-scale *average* path loss in decibel for a distance $d > d_0$ between a transmitter and receiver is

$$\overline{PL}(d) = \overline{PL}(d_0) + 10n \log_{10}\left(\frac{d}{d_0}\right) \qquad (1)$$

where $\overline{PL}(d_0)$—or $\overline{PL}_0$—is the reference path loss, i.e., the average path loss at a reference distance $d_0$. $\overline{PL}_0$ is calculated either by using the formula for free-space path loss (FSPL) [30] or by actual field measurements at $d_0$. Two locations with the same distance $d$ from the transmitter, however, may vary in the amount of surrounding environmental clutter. The corresponding signal levels may therefore differ significantly from the average value. This effect—log-normal shadowing [30]—can be accounted for by adding a zero-mean, Gaussian-distributed random variable $X_\sigma$ with standard deviation $\sigma$

$$PL(d) = \overline{PL}(d_0) + 10n \log_{10}\left(\frac{d}{d_0}\right) + X_\sigma. \qquad (2)$$

Since an attacker is generally more interested in covering the largest possible area than its exact shape, we will ignore log-normal shadowing and thus assume $X_\sigma = 0$.

To estimate the impact of a small-range attack as described in Section IV-A—an attacker targets all STVs in the vicinity—it is important to assess the size of the area controlled by an attacker with a rogue transmitter and low-power amplifier. In general, for a TV to be able to receive a wanted signal in the presence of an unwanted signal, the CCPR of the wanted signal, $\alpha_{\text{wanted}}$, has to be taken into account so that the following condition regarding the received power at the TV's input is met:

$$P_{\text{r,wanted}} \geq P_{\text{r,unwanted}} + \alpha_{\text{wanted}}. \qquad (3)$$

Here, the wanted and unwanted signals are the rogue and regular signal, respectively, and (3) thus becomes

$$P_{\text{t,rogue}} - PL(d_{\text{rogue}}) \geq P_{\text{t,reg}} - PL(d_{\text{reg}}) + \alpha_{\text{rogue}}. \qquad (4)$$

In general, it can be assumed that the power radiated by the regular broadcast station is much higher than that of the rogue station, i.e., $P_{\text{t,reg}} \gg P_{\text{t,rogue}}$. An attacker will therefore have to target areas where the power $P_{\text{r,reg}}$ received from the regular broadcast station is weak, i.e., $P_{\text{r,reg}} \ll P_{\text{t,reg}}$; for reasons of simplification we will assume that $P_{\text{r,reg}}$ is constant in the area under attack. Using (2), the maximum distance $d_{\text{rogue}}$ from the rogue transmitter at which TVs will be able to receive the rogue signal can be calculated as

$$P_{\text{t,rogue}} - PL_0 - 10n \log_{10}\left(\frac{d_{rogue}}{d_0}\right) \geq P_{\text{r,reg}} + \alpha_{\text{rogue}} \qquad (5)$$

$$10n \log_{10}\left(\frac{d_{\text{rogue}}}{d_0}\right) \leq P_{\text{t,rogue}} - P_{\text{r,reg}} - \alpha_{\text{rogue}} - PL_0 \qquad (6)$$

$$d_{\text{rogue}} \leq d_0 \cdot 10^{\frac{P_{\text{t,rogue}} - P_{\text{r,reg}} - \alpha_{\text{rogue}} - PL_0}{10n}}. \qquad (7)$$

For FSPL up to $d_0$, (7) becomes

$$d_{\text{rogue}} \leq d_0 \left(\frac{\lambda}{4\pi d_0}\right)^{\frac{2}{n}} \cdot 10^{\frac{P_{\text{t, rogue}} - P_{\text{r,reg}} - \alpha_{\text{rogue}}}{10n}}. \qquad (8)$$

Finally, the attacker-controlled area is

$$A_{\text{rogue}} = \pi d_{\text{rogue}}^2. \qquad (9)$$

Typical urban and rural radio channels have a PLE ranging from $n = 2.2$ to $4.35$: Measurements conducted in four German cities resulted in an overall mean PLE of $n = 2.7$ [31], for a FSPL reference distance $d_0 = 100$ m. In Santander [32], measurements at a 10 km distance from the transmitter situated on top of a 540 m hill revealed a PLE of 2.17, 2.59, and 2.89 for zones with line of sight (LOS), obstructed by buildings, and obstructed by low hills and buildings, respectively ($d_0 = 1$ m).

*3) Comparison to Previous Work:* The CCPR was ignored in previous work [3], which calculated the attacker's range solely based on FSPL. In its calculation, a target TV would receive the rogue signal under the sole condition that it was stronger than the regular signal, without specifying any margin. We extend this equation to include the required CCPR $\alpha_{\text{rogue}}$, which now gives us the maximum distance $d_{\text{rogue}}$ from the rogue transmitter at which receivers are still able to receive and decode the rogue signal. In addition, we replace FSPL with a generic path loss model that allows for more realistic path loss exponents of $n > 2$.

With the FSPL model ($n = 2$) and no CCPR ($\alpha_{\text{rogue}} = 0$), an amplifier of $P_{\text{t,rogue}} = 1$ W (30 dBm) output power, a center frequency of 500 MHz, and a received regular broadcast signal strength of $P_{\text{r,reg}} = -50$ dBm, an attack range of $d_{\text{rogue}} = 477$ m is calculated in the paper [3]. Based on this radius, the authors state that an attacker can control an area of 1.4 km². This seems to be an error, as calculating the controlled area with $\pi d_{\text{rogue}}^2$ should yield approximately 0.7 km². If we take the required CCPR into consideration, the attack range and hence the controlled area is reduced significantly.

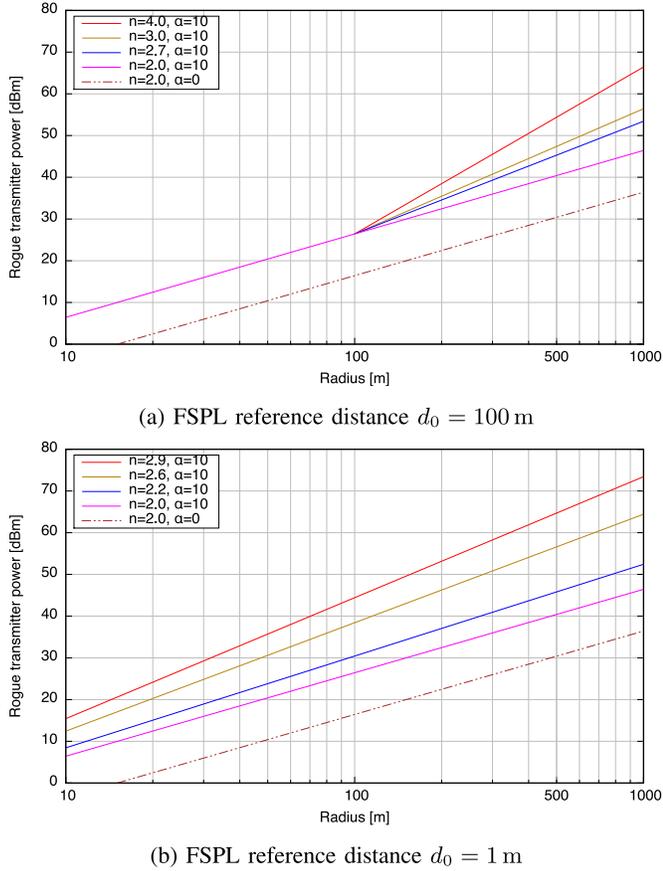

(a) FSPL reference distance $d_0 = 100\,\text{m}$

(b) FSPL reference distance $d_0 = 1\,\text{m}$

Fig. 6. Required rogue transmitter power $P_{\text{t,rogue}}$ as a function of attacker-controlled radius $d_{\text{rogue}}$, CCPR $\alpha_{\text{rogue}}$, and PLE $n$; received regular transmitter power $P_{\text{r,reg}} = -50\,\text{dBm}$, center frequency $f = 500\,\text{MHz}$, and 16-QAM CR 2/3 [12].

A regular broadcast with 16-QAM modulation and a CR of 2/3, by far the most common terrestrial setting in Germany, requires a CCPR of 10 dB according to Table III. Using this CCPR $\alpha_{\text{rogue}} = 10\,\text{dB}$ with (7) yields a maximum attack range $d_{\text{rogue}}$ of 151 m, less than a third of the initially assumed 477 m. The calculation is based on a model that assumes FSPL from the transmitter antenna up to a reference distance $d_0$. From that point on, the signal suffers a path loss with a PLE of $n > 2$. This is illustrated in Figs. 6a and 6b for common reference distances $d_0 = 100\,\text{m}$ and $1\,\text{m}$, respectively.

### B. Mush Zone

An important aspect of the CCPR in the context of terrestrial attacks is the mush zone. It occurs in regions where neither the rogue nor the regular signal are able to achieve the required CCPR. As a result, receivers are unable to lock onto either signal and the TV screen turns dark during an attack. This can be leveraged by broadcasters and authorities to detect commencing attacks and pinpoint the location of rogue transmitters. We call the area, in which receivers are able to lock onto and decode the rogue signal, the attacker-*controlled* area. Adding together this area and the mush area results in the total attacker-*affected* area.

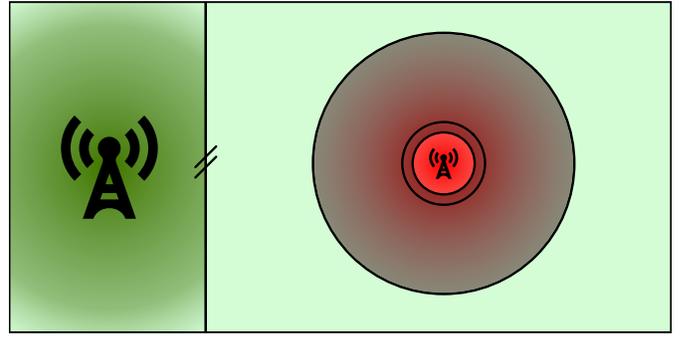

Fig. 7. Relation between rogue and mush zone for differing transmission parameters (16-QAM CR 2/3 vs. QPSK CR 1/2). On the left is the regular broadcast station and on the right the rogue station, which is surrounded by mush zone. The inner shaded annulus is caused by CCPR $\alpha_{\text{rogue}} = 2.5\,\text{dB}$ and the outer by $\alpha_{\text{reg}} = 10\,\text{dB}$. The mush area is over 16 times the size of the attacker-controlled area for these settings [12].

A few simplifications are made. An attacker will target an area that is rather far away from the regular station, due to the significantly smaller power output of the rogue station. At this distance and for the comparatively small area affected by the rogue station, it will be assumed that the signal strength of the regular broadcast is nearly constant in the attacker-affected area and that the attacker uses an omnidirectional antenna. As a result, the shape of the attacker-controlled area and the mush zone is a circle and an annulus, respectively.

In general, the signal strength of the rogue transmitter decreases with an inverse square-law to fourth-law dependence on distance. The mush zone starts where the *rogue* transmitter's signal strength cannot maintain the required CCPR $\alpha_{\text{rogue}}$ w.r.t. the regular signal, and extends to where it has become so weak that the *regular* signal achieves the required CCPR $\alpha_{\text{reg}}$, i.e., $d_{\text{rogue}} < d_{\text{mush}} < d_{\text{reg}}$. If both signals share the same modulation type and code rate, the mush zone corresponds to twice the CCPR. If they use different parameters, this is the sum of both CCPRs (see Fig. 7). The ratio between the radii $d_{\text{rogue}}$ and $d_{\text{reg}}$ of the attacker-controlled and attacker-affected area, respectively, can be calculated using (7) as

$$d_{\text{rogue}} = d_0 \cdot 10^{\frac{-\alpha_{\text{rogue}}}{10n} + c} \tag{10}$$

$$d_{\text{reg}} = d_0 \cdot 10^{\frac{\alpha_{\text{reg}}}{10n} + c} \tag{11}$$

$$\frac{d_{\text{rogue}}}{d_{\text{reg}}} = 10^{-\left(\frac{\alpha_{\text{rogue}} + \alpha_{\text{reg}}}{10n}\right)}. \tag{12}$$

This means that the ratio between the two radii only depends on the corresponding CCPRs and the PLE $n$. The regular broadcast's CCPR $\alpha_{\text{reg}}$ cannot be influenced by the attacker, whereas the rogue signal's CCPR $\alpha_{\text{rogue}}$ can be influenced to some degree as described in Section V-C. Using (12), the

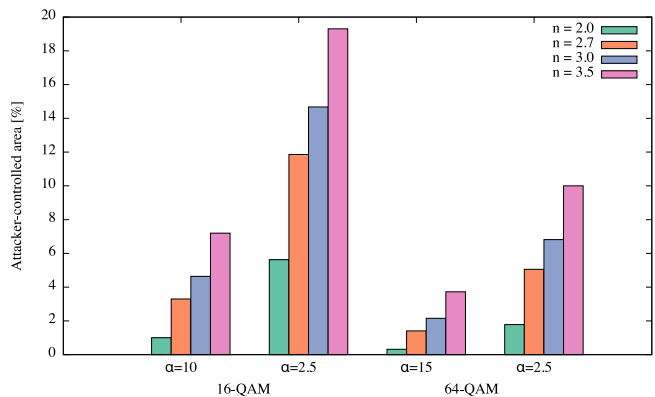

Fig. 8. Percentage of attacker-controlled area out of total attacker-affected area, for various PLEs $n$ ($d_0 = 1$m). Regular station uses 16-QAM and 64-QAM with CR 2/3 ($\alpha_{\text{reg}} = 10$ dB and 15 dB); rogue station uses identical settings ($\alpha_{\text{rogue}} = \alpha_{\text{reg}}$) and the most robust settings, i.e., QPSK with CR 1/2 ($\alpha_{\text{rogue}} = 2.5$ dB) [12].

relation between attacker-controlled and attacker-affected area becomes

$$\frac{A_{\text{controlled}}}{A_{\text{affected}}} = \frac{\pi d_{\text{rogue}}^2}{\pi d_{\text{reg}}^2} = \left(\frac{d_{\text{rogue}}}{d_{\text{reg}}}\right)^2 = 10^{-\left(\frac{\alpha_{\text{rogue}} + \alpha_{\text{reg}}}{5n}\right)}. \quad (13)$$

Overpowering a 16-QAM, CR 2/3 broadcast thus creates a mush area 99 times the size of the attacker-controlled area, assuming an FSPL propagation model ($n = 2$). This is a huge collateral damage and greatly reduces the stealthiness of the untargeted terrestrial attack. Figure 8 shows the percentage of attacker-controlled area for various PLEs $n$ and CCPRs $\alpha_{\text{rogue}}$; the corresponding remaining area is the mush zone.

### C. Variations

An attacker can increase the controlled area by decreasing the required CCPR $\alpha_{\text{rogue}}$, i.e., choose a more robust modulation and code rate. QPSK with CR 1/2 yields the greatest robustness, requiring a CCPR of only approximately 2.5 dB as compared to, e.g., 10 dB for 16-QAM with CR 2/3 (see Table III). The drawback is a reduced available bitrate (factor three for the above example [27]), which cannot accommodate the original transport stream. The attacker either has to decrease the A/V quality by using a higher compression level or remove other elementary streams or programs from the transport stream, both of which increases the risk of being detected.

## VI. ATTACK SCENARIOS

Three different practical attack cases are discussed in this section based on digital terrestrial television coverage simulations on a specific area.

### A. Cases of Attack and Associated Characteristics

The area under analysis is a big urban area of 11km x 15km located in Los Angeles and selected from a larger region that includes the broadcaster's location. One and two story houses prevail in this flat region surrounded by terrain heights that range from a few tens to several hundred meters. In fact, the height of the transmitter considered for the simulations was 1745 meters above sea level. Fig. 9 shows the orography of the terrain as well as the LoS from the transmitter.

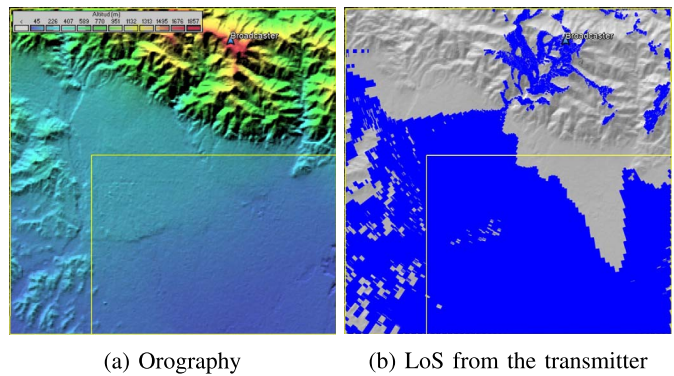

(a) Orography      (b) LoS from the transmitter

Fig. 9. Simulation area.

TABLE V
SIMULATION CASES

| Parameter | Case 1 | Case 2 | Case3 |
|---|---|---|---|
| Reception | Fixed | Indoor Portable | Fixed |
| Broadcaster's Signal | 64-QAM 2/3 | 16-QAM 2/3 | 64-QAM 2/3 |
| Rogue Signal | 64-QAM 2/3 and QPSK 2/3 | 16-QAM 2/3 and QPSK 2/3 | 64-QAM 2/3 |
| Victim | Final user | Final user | Gap filler |

Table V contains a description of each one of the cases under study. They have been proposed to take into account the most influential factors for a successful attack situation.

The cases under analysis differentiate three situations according to the reception mode used by potential victims. There will be users that receive the terrestrial service by means of a rooftop antenna. This case is typical in countries where the terrestrial platform is the dominant delivery method in Europe.

There is a second case where the DTV consumers receive the service indoors. This is a common trend in American and Asian countries.

A third case involves a gap-filler and provides the opportunity for an attacker to target all receivers within the DTV gap-filler coverage area.

The study differentiates several options for the rogue signal. The attacker has two options, one of which is to use the same mode as the broadcaster. This way the rogue signal equals the broadcaster's service capacity. On the negative side (for a potential attacker) the SNR and protection ratio values have to be similar. The second option is to use a more robust mode, requiring a less demanding threshold (SNR and protection ratio) and thus providing a better scenario for the attacker. This option limits the bit rate available to the rogue service (see Section V-C).

Finally it should be highlighted that the users that are most likely to be potential victims (in terms of simplicity from an attacker's perspective) are those that receive the DTV service without employing active amplifiers on their protection infrastructure. If the receiving system is based on amplifiers, the new superimposed signal (rogue signal) on the broadcaster's service is likely to drive the amplifier into the saturation region.

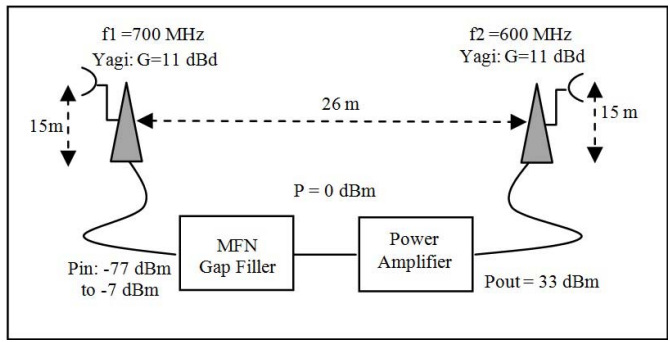

Fig. 10. Gap-filler scheme.

| Param-eter | Description | | DVB-T | |
|---|---|---|---|---|
| | | 64-QAM | 16-QAM | QPSK |
| F | Receiver Noise Figure (dB) | 7 | 7 | 7 |
| Pn | Receiver Noise Input Power (dBW) Pn = F + 10 log(kToB) | -128.16 | -128.16 | -128.16 |
| C/N | Required C/N at Receiver Input (dB) | 15.5[1] | 12.7[2] | 4.8[1]/7.1[2] |
| Ps min | Minimum Receiver Input Power (dBm) | -82.66 | -85.46 | -93.36/ -91.06 |

This will lead to service degradation or loss, thereby decreasing both the number of potential victims and the stealthiness of the attack.

### B. DTV Service Characteristics

The study assumes a DTV service broadcasted by a main transmitter and a multi-frequency network (MFN) gap-filler. The main center transmits the DTV signal from an antenna at 100 m a.g.l. and an effective radiated power (ERP) of 62.6 dBm. A previous network planning exercise was carried out to ensure a signal level of approximately −70 dBm in locations close to the limit of the service area.

In the case of the gap-filler, the receiving subsystem uses an antenna of 11 dBd, at 15 m a.g.l. and a maximum output ERP of 42 dBm. The system accepts a signal input range from −77 dBm to −7 dBm (see Fig. 10).

Table VI describes the DVB-T modes considered in the simulations. These modes are displayed along with the corresponding threshold values. The link budget has been elaborated using the standard DTV network planning calculation procedure as shown in [28] and [33].

### C. Attacker Infrastructure

The attacker uses a reception system in order to acquire the broadcaster's signal and modify the transport stream to be received by the Smart TV without suffering any perceptual change in the audiovisual content. In addition, the attacker uses a transmitting system that will be adapted to each one of the attack cases.

Table VII provides a summary of the reception and transmission parameters considered for simulating the attacks.

---

[1]For a Rice Channel (fixed reception).
[2]For a Rayleigh Static Channel (portable indoor reception).

| | | |
|---|---|---|
| Reception | Antenna | Yagi: 11 dBd |
| | Height (a.g.l) | 5 m if the attacker is at rooftop level |
| | | 2 m if the attacker is in the street |
| Transmission | Antenna | Yagi: 11 dBd or Omni: 0 dBd |
| | ERP | 40 dBm if the attacker uses the yagi antenna |
| | | 29 dBm if the attacker uses the omni antenna |
| | Height (a.g.l) | 5 m if the attacker is at rooftop level |
| | | 2 m if the attacker is in the street |

| | |
|---|---|
| Calculation method | Longley Rice |
| Frequency | 700 MHz |
| Permitivity | 4 (urban ground) |
| Climate | Temperate-continental |
| Location variability | 95 % |
| Surface refractivity | 301 |
| Time variability | 50 % |
| Conductivity | 0.001 S/m (urban ground) |

The geographical location of the attacker with respect to the victim population has also been analyzed in these simulations. There are two possible scenarios for the attacker position. In the first scenario, the attacker is located at an equivalent height similar to the one of the rooftop antennas for DTV reception and the attack can be carried out using either a directive or omnidirectional antenna. A second scenario accounts for an attack carried out using an omnidirectional antenna from a location at street level.

### D. Methodology

The potential success of an attack was studied using field strength maps calculated using the Longley Rice method [34], [35]. Table VIII contains the simulation data considered in this regard.

The basic starting point to evaluate the potential attack was the field strength values of the broadcaster's signal over the area of study. In addition, three other parameters associated to the attacker's signal levels were considered:

1) Attacker's field strength level received at each location within the service area
2) Attacker's interference ratio within the broadcaster's coverage area
3) Broadcaster's interference ratio within the attacker's coverage area

A set of color maps obtained from the values calculated for each one of the previous parameters were processed using a software tool designed specifically for this study. This tool provides the percentages related to the correct reception of the broadcaster's signal (green), the correct reception of the attacker's signal (red), and the mush zone (black) for both, the area of the considered simulation maps and the area of interest for the study (see Fig. 11).

### E. Case 1 Results: Attack Against Fixed Reception Users

The case of an attack against a user that gets the broadcaster's service using a rooftop antenna assumes, for the sake of simplicity and efficiency of the attack, that the victim is located at the fringe of the coverage region. These users receive the broadcaster's signal close to $-70\,\mathrm{dBm}$ and a rogue signal in the range of $-47\,\mathrm{dBm}$. Table IX summarizes the results obtained for this case.

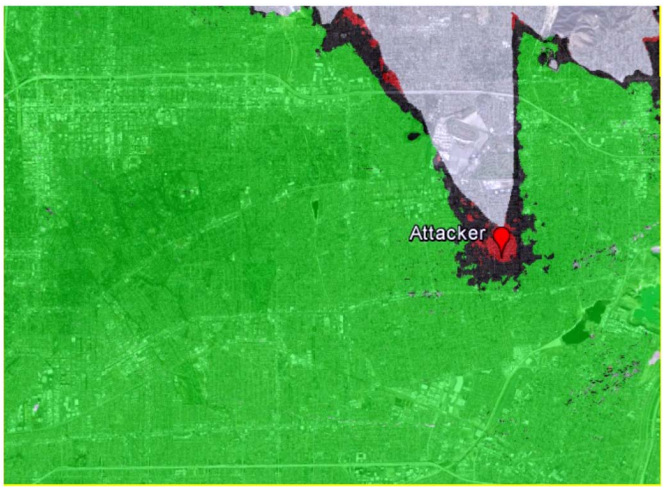

Fig. 11. Example of color map (fixed reception, different DVB-T modes, attacker uses omni antenna at rooftop level).

TABLE IX
LOCATION PERCENTAGE STATISTICS OF AN
ATTACK TO FIXED RECEIVERS

|  | Attacker at rooftop height | | Attacker at street level |
|---|---|---|---|
|  | Directive antenna | Omni antenna | Omni antenna |
| Broadcaster's signal | 96.595% | 95.455% | 97.327% |
| Rogue | 0.005% | 0.021% | 0.004% |
| Mush zone | 3.400% | 4.524% | 2.669% |

The results show that the most critical situation corresponds to the one where the rogue signal is transmitted with an omni-directional antenna from a location slightly higher than the average rooftop height a.g.l. On the weak side for a potential attacker, the mush zone in this case is also the widest one. Nevertheless, the numbers do not provide significant differences between the three scenarios. In the worst situation the ratio between the area where the attack is successful and the mush zone is $0.5\,\mathrm{km}$ / $3\,\mathrm{km}$.

The conditions for the attacker improve if a more robust mode is used (QPSK 2/3). In this case, the success area doubles and the mush zone remains stable (3–4% of the locations of the area of interest).

### F. Case II Results: Attack Against Portable Reception Users

The attack against portable reception users gives significantly worse results, from the attacker's point of view. The number of victims decreases but at the same time the mush zone is also reduced. Table X shows the percentages calculated for the attack against portable users.

According to the values obtained in this case, this type of attack could be chosen to cause harm to a specific group of receivers. The simulations show that the attacker at an equivalent rooftop height would be able to hack an area of $65 \times 65\,\mathrm{m}^2$. The situation does not change significantly if the attack is carried out using a more robust mode.

TABLE X
LOCATION PERCENTAGE STATISTICS OF AN
ATTACK TO PORTABLE RECEIVERS

|  | Attacker at rooftop height | | Attacker at street level |
|---|---|---|---|
|  | Directive antenna | Omni antenna | Omni antenna |
| Broadcaster's signal | 99.8895% | 99.861% | 99.887% |
| Rogue | 0.0006% | 0.003% | 0.002% |
| Mush zone | 0.1099% | 0.136% | 0.111% |

TABLE XI
SIGNAL LEVELS AT THE GAP-FILLER INPUT

| Parameters | Values |
|---|---|
| Broadcaster's signal level received by the attacker ($P_{rba}$) | $-69.6\,\mathrm{dBm}$ |
| Broadcaster's signal level received by the gap-filler ($P_{rbv}$) | $-58.4\,\mathrm{dBm}$ |
| Attacker's signal level received by the gap-filler ($P_{rav}$) | $-38.8\,\mathrm{dBm}$ |
| $P_{rav} - P_{rbv}$ | $19.6\,\mathrm{dB}$ |

TABLE XII
VICTIM LOCATION PERCENTAGES OF INTERFERENCE (GAP-FILLER)

| Parameter | Values |
|---|---|
| Broadcaster's coverage ($f_1$) | 91.0293% |
| Rogue (received directly at $f_1$) | 0.0005% |
| Mush zone | 8.9702% |

### G. Case III Results: Attack Against a Gap-Filler

This case describes a situation where the number of potential victims is relatively large. The gap-filler is attacked by overpowering the input signal with the rogue component, which, in consequence, will be re-broadcasted by the output power modules of the gap-filler over its service area. It was assumed that the attacker system location to send the rogue component using a directional antenna is at a distance of $30\,\mathrm{m}$ away from the gap-filler site. In this case, a major problem to be solved by the attacker is the adjustment of appropriate input levels at the gap-filler receiving antenna, avoiding transmitter saturation. Table XI shows the signal values assumed by this simulation exercise.

These values lead to conclude that the attack would be successful, since the rogue signal would be received 20 dB over the broadcaster's service; a value high enough for correct reception of the mode 64-QAM 2/3. In this scenario, vulnerable Smart TVs within the gap-filler service area will be compromised.

Also, the rogue signal can interfere with the broadcaster's signal received by final users at the gap-filler input frequency (f1 in Fig. 10). Table XII provides data associated to this case. These numbers show that the percentage of locations where this situation occurs is close to 9% of the simulation area.

## H. Conclusion

The main conclusion of the simulated cases is that DTV consumers that use vulnerable Smart TVs can be attacked by using an infrastructure that is relatively simple. A second relevant conclusion is that the number of potential victims is quite limited (percentages well below 1% of the area under study). The worst case is an attack to users receiving the broadcaster's DTV service by means of a rooftop antenna.

Nevertheless, in this case the mush zone is wider and thus the stealthiness of the attack is compromised. From an attacker's perspective the success is higher when an omnidirectional transmitting antenna is located over the rooftop height. However, the attack at street level is also interesting since the mush zone will be reduced, especially if the victims use a rooftop antenna.

Additionally, the mode of the rogue signal is not critical for a successful attack, but has certain effect in the total number of victims, since in general is higher when the attacker uses a more robust mode.

Another remarkable aspect is the variability of the attack situations. The size of the regions where the attacker has success comprises distances from 50 m to 1 km, depending on the characteristics of the area, the features of the attacking system as well as the receiving system used by the victims. This variability is also observed when analyzing the mush zone, which ranges from hundreds of meters to several kilometers. Finally, the attack against a gap-filler represents a singular case, where the number of victims can be significantly higher — provided that the input levels at the receiving antenna of the gap-filler are properly adjusted.

## VII. Field Measurements

Broadcasters might have an interest in determining potential attack ranges in practice. This would typically involve specific locations within an area serviced by existing regular broadcast or relay stations. We propose a method that can be used in existing single-frequency networks (SFN), does not interrupt service, and only requires an add-on license to operate a low-power on-channel gap filler.

### A. Methodology

A common method for analysis of an SFN is based on channel estimation data collected from the scattered and continuous pilots of a received DVB-T signal. The resulting transfer function of the broadcast channel can then be transformed to obtain the channel impulse response. Peaks represent the transmitters in the SFN and the delays between them correspond to the distance from the receiver, as depicted in Fig. 12.

An attack can be simulated by operating a gap filler that is synchronized to the SFN. Field measurements in the vicinity of the gap filler will then reveal the received power of each signal source via the channel impulse response. If the signal from the gap filler exceeds all other signals by the required amount of CCPR, this location would be part of the attacker-controlled area in a real attack. The attacker-affected area can be determined similarly (see Section V-B).

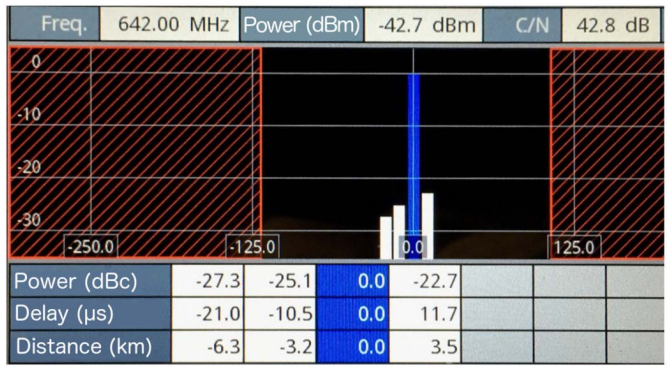

Fig. 12. The first two bars represent the DVB-T2 signals from the radio towers at Alexanderplatz and Schäferberg, respectively, and their respective echoes generated by the on-site on-channel gap-filler. The Alexanderplatz echo is the strongest signal with a margin of more than 20 dB.

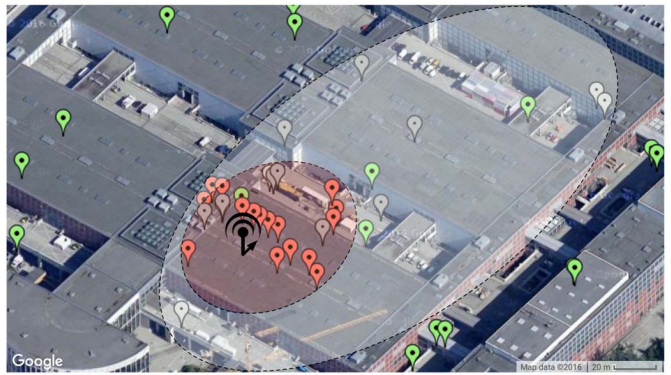

Fig. 13. Exemplary approximation of attacker-controlled (red) and affected (gray) area (mush zone). Red and green markers correspond to measurement points with service controlled by the rogue or regular station, respectively; dark and light gray markers indicate mush zone, where the rogue or regular signal is stronger, respectively.[3]

### B. Field Trial

We performed a field trial in Berlin, Germany, to evaluate this method in practice. Two radio towers, located at Alexanderplatz and Schäferberg, provide the city with the recently launched DVB-T2 service. They operate as an SFN on channel 42, each with 50 kW ERP, using DVB-T2 variant G8 [29]. To improve the DVB-T2 signal coverage within the International Congress Centrum (ICC) building hosting the International Radio Exhibition (IFA) 2016, an additional repeater is operated for the duration of the exhibition. The repeater is fed with a signal received by a rooftop antenna facing Alexanderplatz, which is retransmitted on-channel with an output power of 300 mW inside a hall of the building. Figure 13 shows the location and orientation of the transmitting antenna (Kathrein 75010128).

A series of channel impulse response measurements were acquired in the vicinity of the repeater. Figure 12 shows one of these measurements: Each signal path is represented as a bar in a diagram, where the x- and y-axis indicate the delay and strength relative to the strongest signal, respectively. Here, the first signal to arrive originates from Alexanderplatz, followed by Schäferberg, and the respective echoes from the repeater; we are only interested in the Alexanderplatz signal and its echo. This data was acquired close to the repeater and therefore the repeater signal is much stronger than the original signal.

We then proceed to plot this measured difference in signal power on a map, for the entire measurement series. Depending on this value, locations are categorized as controlled by a potential attacker (red), mush zone (gray), or regular service (green). To qualify for the first or last category, the echo signal has to exceed or fall below the original signal by more than its CCPR, respectively; anything in between results in mush zone. The results of our field trial are presented in Figure 13.

## VIII. CONCLUSION

This work demonstrates that DTV broadcasts can be abused to permanently and stealthily infect Smart TVs with malware, in less than ten seconds. It also reveals, however, that broadcast-based attacks are subject to serious limitations in terms of range and stealthiness. We thus revise previously published calculations regarding the size of an attacker-controlled area down by a factor of twenty, for a typical broadcast scenario. Furthermore, we show that such an attack creates a mush area without service coverage, which significantly aids broadcasters and authorities to detect and locate a commencing attack. Broadcasters should take appropriate measures especially to protect neuralgic points such as terrestrial gap fillers and cable TV headends. In the long term, broadcast standards should include mandatory authentication information, such that receivers can securely verify the origin and integrity of digital broadcasts.

## ACKNOWLEDGMENT

The authors would like to thank Andrew Karpow, Alexandra Mikityuk, and the Helmholtz Research School for Security Technologies for their help and support with this work.